\documentstyle[12pt,epsfig,amsmath]{article}  
\def\M{M_{K\overline K}}   
\def\KK{K\overline K}   
\title{{\bf  ${\bf K{\overline K}}$ photoproduction and ${\bf S-P}$ wave interference}}   
\author{\vspace{0.4cm}{\bf \L{}ukasz Bibrzycki$\dag$, Leonard Le\'sniak$\dag$, Adam P. Szczepaniak$\ddag$} \\   
\small {$\dag$ Department of Theoretical Physics, }\\   
\small { The Henryk Niewodnicza\'nski Institute of Nuclear Physics,}\\
\small { Polish Academy of Sciences, PL 31-342 Krak\'ow, Poland}\\      
\small{$\ddag$ Physics Department and Nuclear Theory Center, }\\   
\small{ Indiana University, Bloomington, IN 47405, USA} }    
    
\begin{document}    
     
\maketitle     
      
\begin{abstract}   
   
 Results of a new analysis of the $K^+K^-$   
 photoproduction at two photon energies $E_\gamma = 4\mbox{ GeV}$ and    
 $5.65\mbox{ GeV}$ with a particular emphasis on the $S$-wave  
 production are presented. We show that the proper treatment of all the helicity  
 components of the $S$- and $P$-waves enables one to eliminate the  
reported discrepancies in the extraction of the $S$-wave  photoproduction cross   
section from experimental data.    
   
\end{abstract}    
   
\maketitle   
   
\section{Introduction}   
    
\hspace{0.5cm} From the early days of QCD, light meson spectroscopy played an   
 important role in  development of the theory and in    
 understanding of its low energy structure.    
  The flavour symmetry of QCD originates in part from the observed $SU(3)$    
  multiplet structure  of the light pseudoscalar and vector mesons. More     
  recently, scalar and tensor spectra have provided evidence for a    
  possible over-population of the $Q{\overline Q}$ spectrum and for the existence of    
 gluon-rich states \mbox{[1-4]}.  
  The possible existence of    
  gluonic excitations  is  one of the most intriguing features of the   
 meson spectroscopy. There is tantalizing evidence of exotic,    
 hybrid mesons in the spectrum around $1.6\mbox{  
  GeV}$ [5-7], and future   
 experiments proposed for JLab and GSI in the light and charm sector   
 respectively will map out the exotic spectrum. The glueball signature   
 comes primarily from the analysis of the Crystal Barrel $p{\bar p}$   
 and WA102 central production data~\cite{CB,WA102}. Detailed mapping of   
 various scalar meson decay channels led to the identification of   
 $f_0(1370)$, $f_0(1500)$ and $f_0(1710)$ states, all expected to contain   
 significant gluon components.    
  While the genuine QCD resonances in the scalar channel are not   
 expected to occur below 1 GeV, the low energy region can be   
 studied using standard, low energy expansion techniques.  These include    
 the effective range expansion, $N/D$ and other methods  based on    
 analyticity coupled with truncation of the unitarity condition.    
 The parameters of the    
  soft meson-meson interactions  {\it e.g.} subtraction constants,   
 form factors, coupling strengths, {\it etc.} effectively correspond to    
 local potentials smeared over distance scales of the orders of   
 $\sim 1~ \mbox{fm}$, {\it i.e.} close to the pion root mean square radii.    
 The strength of  the interaction can be constrained from chiral   
 symmetry or simply fitted to the data [8-11].    
 As a result one obtains a very good description of the    
 spectrum including the resonance region. Furthermore, the behavior    
  of the scattering amplitude in the complex energy plane enables one to   
  establish the existence of dynamical resonances such as    
 the isoscalar $\sigma(600)$ and $f_0(980)$ and the isovector  
 $a_0(980)$ mesons. The last two are particularly  
  interesting as they are very sensitive to the details of the   
 meson-meson interactions. This is due to the   
 proximity of the $\KK$ threshold. In particular the detailed   
 structure of the   $f_0(980)$ {\it e.g.} whether it is a genuine   
 bound state or a virtual bound sate, strongly   
 depends on the threshold $\KK$ interaction parameters. Since the $\KK$ channel  
   also influences the higher mass region a proper description of its   
 dynamics is crucial for a partial wave analysis and the identification of   
 exotic, hybrid or glueball signals.    
 Furthermore the $\KK$ system is relevant for testing the origins of CP   
 violation and possible signals of CPT violation via hadronic   
 interferometry~\cite{CPT}.    
    
 The data on the near threshold $\KK$ production are very   
 scarce and come mainly from high-energy peripheral   
 production~\cite{KK,omega}.   
 In the medium-energy region, $E^{lab}_{\gamma}\sim\mbox{a few GeV}$,   
  it is advantageous to study the $\KK$ system in photoproduction. A real   
 photon couples strongly to vector mesons and near the $\KK$   
 threshold $\phi$ photoproduction   
   dominates the $\KK$ spectrum. In this energy range the   
 $\phi$ photoproduction cross section is large,   
 ~$\sigma(\gamma P \to \phi p) \sim 0.5~ \mu b$, and has an energy dependence   
 characteristic for diffraction.    
 The $s$-channel baryon resonance production cross section is highly   
 suppressed  at   
 these energies, and the $t$-channel meson exchange ($\pi$) is marginally   
 relevant only at $E_\gamma$ very close to threshold~\cite{Lee}.    
 Finally, the $S$-wave $\KK$ state, dominated in this mass range by the   
 $f_0(980)$ in the isoscalar and by the $a_0(980)$ in the isovector  
  channel,   
  can be accessed via interference with the   
 $\phi$ meson in the $P$-wave .    
 This $S-P$ interference  was first explored in   
 experiments at DESY~\cite{Fries,Behrend} and Daresbury~\cite{Barber},    
 and could be further studied with high statistics at an energy    
 upgraded JLab.

 In the  analysis of the data from the DESY and Daresbury experiments,    
 however, the rich spectrum of the $S$-wave $\KK$ has not been   
 fully  explored. In both cases the  $S$-wave  was parameterized as   
 either a simple Breit-Wigner resonance or a nonresonant background.    
  The two analyses yielded results for the $S$-wave production cross section  
  different from    
 each other by more than one order of magnitude. Finally, apart of the $K^+K^-$  
 mass spectrum, only one moment $\langle Y^1_{\;0}\rangle$, describing   
 the $K^+$ angular distribution, has  been analyzed and very simplified   
 assumptions about the nucleon spin dependence have been done.  
   
  In this paper we use the results of a recent calculation of the coupled   
 channel scalar-isoscalar and scalar-isovector spectra together with a diffractive model   
 for $P$-wave production to interpret the existing $S-P$ interference   
 data and estimate the $S$-wave production cross section with a better accuracy than in the  
 phenomenological analyses already published in [16-18].  
 Here we shall discuss a complete set of six moments $\langle Y^L_{\;M}\rangle$  
 of spherical harmonics including $L=0$ and $L=1$.  
   In the following section we shall   
 present the theoretical foundation of the $S$- and $P$-wave   
 photoproduction and discuss the main features of the existing data.    
 In Sect. 3 we present results of the numerical analysis and    
  fits to the data. Conclusions and outlook are given in Sect. 4.    
   
\section{${\bf S}$- and ${\bf P}$-wave  photoproduction}   
   
\hspace{0.5cm}In this paper we consider unpolarized $\KK$ photoproduction reaction    
\begin{equation}   
\gamma p \to K^+  K^- p,    
 \end{equation}   
 for incident photon laboratory energies $E_\gamma$ of the order of a few   
 GeV. This is an optimum energy range for the $S$-wave $\KK$   
 production. In this energy range the process is dominated by the    
  pomeron exchange, leading to $\phi$-meson production which becomes   
 even more    
 important as the photon energy increases.  However, $t$-channel   
 processes expected to be responsible for the \mbox{$S$-wave} production   
 decrease rapidly with energy.    
 The experimental evidence for the $S-P$ interference in the $E_\gamma$ range    
 between 2.8 and 6.7 GeV was presented in [16-18].  
The values of the $S$-wave photoproduction cross sections found in these   
 experimental analyses varied between 2.7 and 96 nb.   
 In both cases, in the effective mass range  
 $1.00 \mbox{ GeV} < \M < 1.04 \mbox{ GeV}$, in the  rest frame of the $\KK$ system an asymmetry in the kaon polar   
 angle distribution was observed. This can only be the case if there    
 are odd powers of  $\cos\theta$  in the angular distribution.    
 For low partial waves this implies presence of both $S$- and   
 $P$- waves. This feature of the angular   
 distribution is independent of the magnetic-quantum number {\it i.e.}   
 choice of the quantization  axis. Due to the empirical $s$-channel   
 helicity conservation, diffractive production is most naturally   
 analyzed in the $s-channel$ helicity frame, which in our case    
 corresponds to the  rest frame of the $\KK$ pair with the $z$-axis    
 anti-parallel to the direction  of the recoiling  nucleon.   
In another reference frame, called the $t$-channel frame or the   
Gottfried-Jackson frame, the $z$-axis is chosen along the direction of the   
photon beam with the $\KK$ pair at rest~\cite{Jack}.  
  In both cases the $y$-axis is    
 perpendicular to the production plane.   
  
As discussed in   
 ~\cite{Ji},  in this energy range, the $S$-wave production is   
  expected to be dominated  by vector $\rho$, $\omega$ and   
 pseudoscalar $\pi$, $K$ $t$-channel exchanges.  
    
 As a function of the momentum transfer squared $t$, the kaon   
 pair invariant mass $\M$ and the $K^+$ decay angles $\Omega =   
(\theta,\phi)$ in the $s$-channel frame    
 the photoproduction amplitude restricted  to $S$- and $P$-wave s can be   
 written as    
\begin{equation}    
T_{\lambda_\gamma \lambda \lambda'}(t,\M,\Omega) = \sum_{L=S,P;M}   
T^L_{\lambda_\gamma \lambda \lambda' M}(t,\M)~ Y^L_{M}(\Omega).   
\end{equation}   
Here $\lambda_\gamma$, $\lambda$ and $\lambda'$ denote the photon,   
 target proton and recoil proton helicities, respectively, and $Y^L_{M}(\Omega)$  
 are the spherical harmonics. The corresponding    
 four-momenta will be denoted by $q$, $p$, $p'$, and $k_1$ and $k_2$   
 will be used for the $K^+$ and the $K^-$ momenta, respectively.    
 The  unpolarized differential cross section is given by   
\begin{equation}   
{{d\sigma}\over {dt~d\M~ d\Omega}} =  
\frac{\kappa_f }{32~ (2\pi)^3~ m_p^2~ E_\gamma^2}  
 {1\over 4} \sum_{\lambda_\gamma\lambda\lambda'}       \label{ds/dtdmdo}  
|T_{\lambda_\gamma \lambda \lambda'}(t,\M,\Omega)|^2,    
\end{equation}  
where $m_p$ is the proton mass, $m_K$ is the kaon mass and   
$\kappa_f= \sqrt{ {\M^2\over 4} - m_K^2}$ is the kaon momentum in the rest  
system of the $\KK$ pair.  
The $S$-wave amplitude $T^S$ and the $P$-wave amplitude $T^P$ have   
been described in ~\cite{Ji,acta}, respectively. Here   
 we will briefly summarize the basic properties of these amplitudes.    
 The $S$-wave  $\KK$ production is parameterized as a double    
  \mbox{$t$-channel} exchange.   
In the upper meson vertex we use a simple meson exchange and allow for an  
interaction of the two produced mesons in the final state.   
 The dominant exchanges for the $S$-wave $\KK$ production are shown in   
 \mbox{Fig. 1}.      
At the nucleon vertex we use either normal or   
 Regge propagators of the exchanged vector mesons. The normal   
propagator of the   
 vector meson of the mass $m_e$ is equal to   
 $ (t-m_e^2)^{-1}$ and the Regge type propagator reads  
\begin{equation}   
 -[1 - e^{-i\pi\alpha(t)}]~\Gamma(1 - \alpha(t))~    
(\alpha's)^{\alpha(t)}/(2s^{\alpha_0}),   
\end{equation}   
where we have the vector meson trajectory $\alpha(t) = \alpha_0 +   
\alpha'(t-m_e^2)$, $\alpha_0 =1$ and  $\alpha'= 0.9$ GeV$^{-2}$.    
\begin{figure}[htbp]  
\begin{center}     
\mbox{\epsfxsize 8cm\epsfysize 5cm\epsfbox{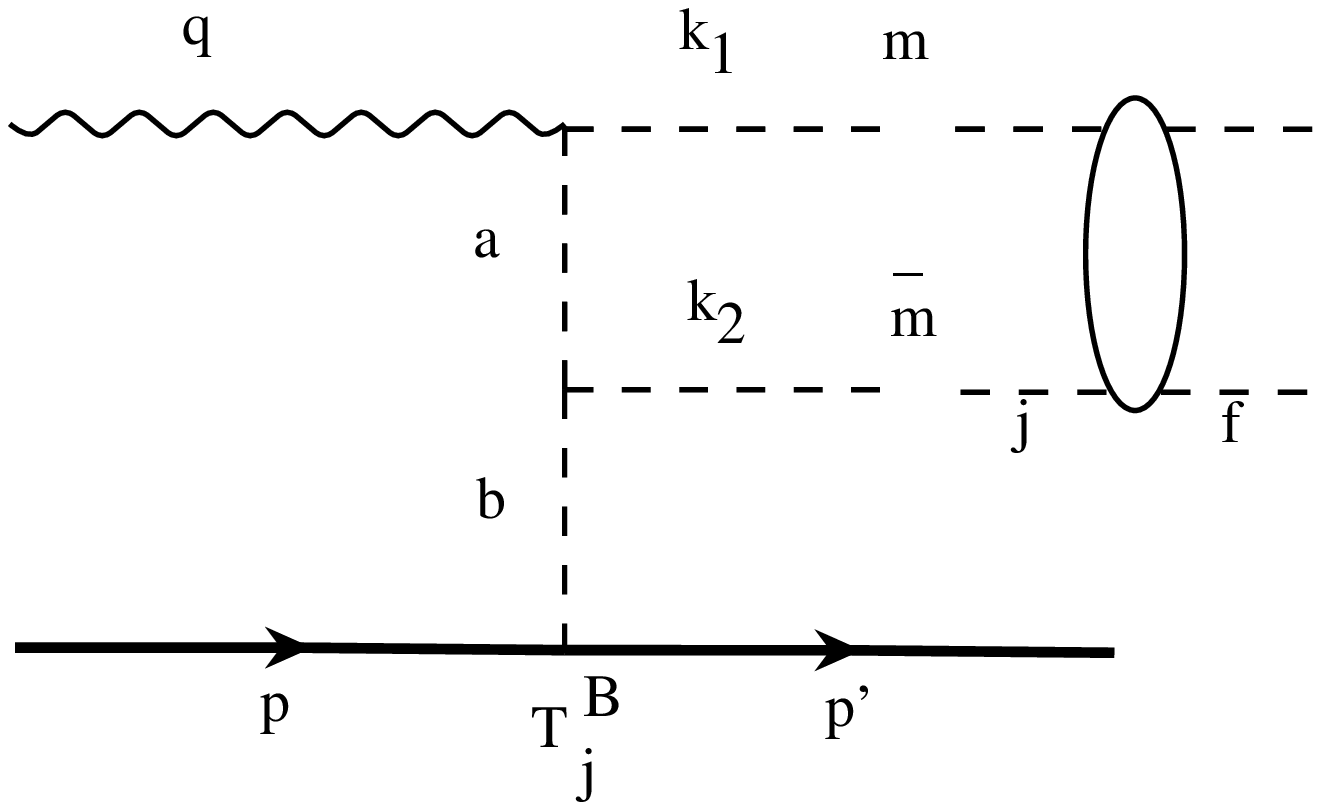}}  
\end{center}     
{Fig. 1: The amplitude for the $S$-wave  $\KK$   
   production. The $a$ in the Born amplitude $T^{B}_{j}$   
   stands for $\pi$, $K$, $\rho$ and $\omega$ mesons, the $b$ stands for  
   $\rho$ or $\omega$ mesons and $m{\overline m}$ denotes   
   either a $\pi\pi$ or $\KK$ pair. The oval represents the   
   final state rescattering amplitude.}  
\end{figure}          
The final state interactions include the $\pi\pi$ and   
 $\KK$ channels. Both interactions  
can have a resonant character. This is important since the \mbox{$S$-wave } in the   
 mass region of  interest, $\M \sim 1\mbox{ GeV}$, is dominated by the    
 $f_0(980)$ and $a_0(980)$ resonances. We should notice that the $K^+K^-$  
 system is an equal mixture of the isospin 0 and isospin 1 states. The isoscalar 
  $f_0(980)$ resonance has a 
  main branching to $\pi\pi$ and in addition an important coupling to $\KK$. The 
isovector $a_0(980)$ resonance lies also very close to the $\KK$ threshold so one  
should take it into account in the calculations of the final state interactions.   
 The $S$-wave $K^+K^-$ photoproduction amplitude $T^S_f$ can  therefore   
 be  written as    
\begin{equation}   
T^S_{\lambda_\gamma \lambda \lambda'} =   
 {\bar u}(p',\lambda') J^S_\mu(p',p,q,\M) \epsilon^\mu (q,\lambda_\gamma)   
 u(p,\lambda)   
\end{equation}  
and decomposed as $T^S_f=\frac{1}{2}[T^S_f(I=0)+T^S_f(I=1)]$.  
 Here $\epsilon$ is the photon helicity four-vector and    
 $J^{S}_{\mu}$ is the $S$-wave projection of the appropriate current operator. As   
 illustrated in Fig. 1  
 each $S$-wave amplitude is a sum of products  
 of the Born amplitudes $T^B_j(I)$ describing the    
 $t$-channel meson or Regge exchanges and the final state   
 interaction factors $F_{jf}(I):$   
\begin{equation}  
T^S_f(I)=\sum_{j=\pi\pi,\KK} T^B_j(I) F_{jf}(I).      \label{TI}  
\end{equation}   
If we restrict ourselves to the on-shell part of the final state coupled  
channel interactions, represented by the 2$\times$2 $S$-matrix elements 
$S^I_{jf}$, then the factors $F_{jf}(I)$ can be written as  
\begin{equation}  
F_{jf}(I)={{1}\over{2}} (\delta_{jf} + S^I_{jf})\sqrt{\frac{\kappa_j}{\kappa_f}},  
                                                       \label{Fjf}     
\end{equation}  
where $\kappa_j=\sqrt{{M_{m{\overline m}}^2\over 4} - m_j^2}$ and 
 $M_{m{\overline m}}$ is the effective mass of the $m{\overline m}$ pion or kaon  
 pair and $m_j$ is the pion or kaon mass.  
The explicit forms of the $S$-wave Born amplitudes $T^B_j$ are   
 given in ~\cite{Ji}.  
The magnitude of the off-shell part of the final state interaction amplitude   
is much less certain than the on-shell part as has been shown in ~\cite{Ji}.  
Thus we prefer here to use only the on-shell part of the amplitude and   
correct it later by a constant modification factor.   
In comparison with ~\cite{Ji} we have added in these amplitudes a   
 $t$- dependent factor $exp(B_S t)$, where the parameter  
 $B_S=1.07$ GeV$^{-2}$ is responsible for a spatial dimension of the meson   
(kaon or pion) coupled to the photon. In ~\cite{Ji} the meson in the upper   
 vertex in Fig. 1 was treated as a pointlike particle.     
 The isoscalar $S^I_{jf}$ matrix elements can be expressed by the   
 scalar- isoscalar phase shifts $\delta$ and inelasticity $\eta$ in two channels $\pi\pi$  
and $\KK$:  
 \begin{equation}   
S^{I=0}_{jf} = \left( \begin{array}{cc}    
  \eta e^{2i\delta_{\pi\pi}^{I=0}} & i\sqrt{1 - \eta^2}   
  e^{i(\delta_{\pi\pi}^{I=0} + \delta_{\KK}^{I=0})} \\   
 i\sqrt{1 - \eta^2}   
  e^{i(\delta_{\pi\pi}^{I=0} + \delta_{\KK}^{I=0})}&     
  \eta e^{2i\delta_{\KK}^{I=0}}   
 \end{array} \right).                                    \label{SI0}  
\end{equation}  
 These elements have been computed in ~\cite{KLL} and here we use   
 the solution A corresponding to the so-called "down-flat" data of the phase   
 shift analysis ~\cite{KLR}.    
  
The I=1 $\KK$ interaction near 1 GeV is very strongly influenced by the     
$a_0(980)$ resonance which decays dominantly into the $\pi\eta$. A coupled  
channel model including the $\pi\eta$ and $\KK$ states has been    
formulated in ~\cite{LL} and recently compared to the existing data  
~\cite{Furman}. As a result one   
obtains the $\KK$ scalar-isovector $S$-matrix elements $S^{I=1}_{jf}$ which  
 differ from the scalar-isoscalar $S$-matrix elements written in (\ref{SI0}).  
This new information has been incorporated into calculations of the final   
state $K^+K^-$ interactions.  
    
The $P$-wave amplitudes corresponding to a given projection M of the $K^+K^-$  
angular momentum on the quantization axis can be handled in a similar way  
as the $S$-wave , {\it i.e.} we write     
\begin{equation}   
T^P_{\lambda_\gamma \lambda \lambda' M} =   
 {\bar u}(p',\lambda')~J^P_{{\mu}M} (p,p',q,\M)    
 ~\epsilon^\mu(q,\lambda_\gamma)~ u(p,\lambda),    
\end{equation}   
with $J^P_{{\mu}M}$ being the $P$-wave projection of the current.  
The $P$-wave current is described in detail in ~\cite{acta}. In particular,    
 it is saturated by the diffractive $\phi$-meson production,    
 \begin{equation}   
J_\mu  = {{i~F(t)}\over {M^2_\phi - \M^2 - iM_\phi \Gamma_\phi}}   
\left[ \gamma^\nu q_\nu (k_1 - k_2)^\mu - q^\nu (k_1 - k_2)_\nu   
  \gamma^\mu \right],                                   \label{p}   
\end{equation}   
where $\gamma^\mu$ are the Dirac matrices, and $M_\phi$ and $\Gamma_\phi$ are $\phi$ mass and width, respectively.  
The Lorentz-Dirac structure of the current is motivated    
 by the Donnachie-Landshoff model for the pomeron exchange which   
 assumes vector coupling of the Po\-me\-ron to hadrons~\cite{DL}.    
 The two terms in ~(\ref{p}) are needed to preserve electromagnetic current   
 conservation, $q^\mu J_\mu = 0$.  
One can show that $q^\mu J^P_{{\mu}M}= 0$ is satisfied independently by each  
  projection of the $\KK$ spin, $M=\pm 1,0$.   
 Then in the phenomenological analysis of the data one  
 can separately modify different $J^P_{\mu M}$ components by multiplying them  
  by constant factors.  
  The function $F(t)$ is a   
  phenomenological function which will be suitably parameterized to   
 reproduce, at fixed energy,  the $t$-dependence of the observed  
 \mbox{$\phi$   
  photoproduction.} Its analytical form will be specified in the next section.  
 We note, however, that this model for the $P$-wave  does not lead to   
 significant suppression of the single helicity flip amplitude.    
 In particular in the high energy limit of   
 $s \simeq ~ 2E_\gamma m_p \gg |t|$ and $s \gg \M$ the    
 helicity non-flip, single-flip and double-flip amplitudes with    
 $\lambda_\gamma=1$ behave as    
   
\begin{equation}   
T_{1\lambda' \lambda1} \propto 2 N_{\lambda'\lambda} \M, ~~~    
T_{1\lambda'\lambda0} \propto N_{\lambda'\lambda} \sqrt{-2t},  ~~~   
T_{1\lambda'\lambda-1}  \to O(1/s),    
\end{equation}   
respectively. The proportionality constant $N_{\lambda'\lambda}$ contains the    
 Breit-Wigner propagator of ~(\ref{p}) and is finite as  
 $t'=t-t_{min} \to 0$. The   
 diagonal elements of the matrix $N$ are finite and the   
 off-diagonal ones,  corresponding to nucleon helicity flip, are $O(1/s)$. Thus    
  the only suppression of the photon-meson coupling comes from the   
 angular momentum conservation factor $t^{|\lambda_\gamma-M|/2}$, but   
 it is otherwise finite at high energy. The existing data on $\phi$   
 photoproduction suggest that the helicity flip amplitude is of the   
 order of $10\%$ of the dominant, helicity non-flip one at photon   
 energies below $10 \mbox{ GeV}$~\cite{Barber}.    
 Thus, qualitatively  the model has   
 the correct features but the quantitative  agreement may require    
 adjusting the normalization and the phase of the helicity flip amplitudes.    
This can be done by multiplication of these amplitudes by a constant complex  
factor $C_{10}~exp(i\phi_{10})$, where  $C_{10}$ is a real positive number  
and $\phi_{10}$ is an additional phase of the $P$-wave   amplitude with $M=0$.  
We will not modify phases of the dominant helicity non-flip amplitudes,  
however we do change slightly their moduli as well as the moduli of small  
double helicity flip amplitudes to keep the total $P$-wave   cross-section  
untouched. The phases of the small double helicity flip amplitudes are the 
 same as the phases of the corresponding non-flip amplitudes. Similarly, for the $S$-wave   amplitudes we introduce a constant   
modification factor equal to $C_{00}~exp(i\phi_{00})$. It is assumed  
that the parameters $C_{10},  
\phi_{10}, C_{00}$ and $\phi_{00}$ do not depend on proton helicities   
$\lambda$ nor $\lambda'$.  
The experimental data on the angular distribution of the $\KK$ pair are    
 given in terms of the moments $\langle Y^{L}_{M}\rangle$  of the angular    
 distribution evaluated in the $s$-channel helicity frame,   
\begin{equation}  
\langle Y^{L}_{M}\rangle     
  \equiv \int_{t_1}^{t_2}\!\! dt \int \!d\Omega~ Y^{L}_{M}(\Omega)   
  {{d\sigma}\over {dt~d\M~ d\Omega}}  =   
 {\cal N} \!\int\! \!\int \!dt~d\Omega \sum_{\lambda_\gamma\lambda\lambda'}   
  |T_{\lambda_\gamma\lambda\lambda'}|^2 Y^{L}_{M}(\Omega),    
\end{equation}   
 where $\cal N$ takes into account the photon flux and the 1/4 factor standing   
 before the sum in (\ref{ds/dtdmdo}). The lowest moment is normalized to the  
 $K^+K^-$ mass distribution integrated over   
 the momentum transfer squared range limited by $t_1$ and $t_2$:    
\begin{equation}   
\langle Y^{0}_{\;0} \rangle = {1\over {\sqrt{4\pi}}}{d\sigma \over d\M}.    
                                                       \label {Y00}   
\end{equation}   
In terms of the $S$- and $P$-partial waves, the non-vanishing    
 moments are \mbox{given by}   
\begin{equation}  
\begin{split}  
&\langle Y^{0}_{\;0} \rangle = {{\cal N} \over {\sqrt{4\pi}}}  
\left( |S|^2 + |P^2_+| + |P^2_-| + |P^2_0| \right),\\  
&\langle Y^{1}_{\;0} \rangle =  {{\cal N}  \over {\sqrt{4\pi}}}  
\left( S P^*_0  + S^* P_0 \right),\quad  
\langle Y^{1}_{\;1} \rangle = {{\cal N}  \over {\sqrt{4\pi}}}  
\left(  P_+ S^*  - S P_-^* \right),\\  
&\langle Y^{2}_{\;0} \rangle = {{\cal N}  \over {\sqrt{4\pi}}} \sqrt{1\over 5}  
\left( 2 P_0 P_0^* - P_+ P^*_+ - P_- P^*_-  \right),\\  
&\langle Y^{2}_{\;1} \rangle = {{\cal N}  \over {\sqrt{4\pi}}}  
\sqrt{3\over 5} \left( P_+ P_0^* - P_0 P_-^*  \right),\quad  
\langle Y^{2}_{\;2} \rangle = {{\cal N}  \over {\sqrt{4\pi}}} \sqrt{6\over 5}  
\left( - P_+ P_-^*\right).  
\end{split}  
\end{equation}  
Here $S$, $P$ stand for $T^S$ and $T^P$ amplitudes, respectively, and summation  
over photon and nucleon spin indices is implicit, {\it e.g.},   
\begin{equation}   
P_+ P^*_0 = \sum_{\lambda_\gamma\lambda\lambda'}   
T^P_{\lambda_\gamma \lambda \lambda' 1}    
 ~T^{*P}_{\lambda_\gamma \lambda \lambda' 0}.    
\end{equation}   
The dominant $P_+$ and $P_-$ waves originating from   
$M=\lambda_\gamma=+1$ and \mbox{$M=\lambda_\gamma=-1$} helicity non-flip   
 production will manifest themselves in a large, positive   
 $\langle Y^0_{\;0} \rangle$ and a large negative    
 $\langle Y^2_{\;0} \rangle$ near $\M = 1.02\mbox{ GeV}$ -- the   
 mass of the $\phi$ resonance. This is indeed the dominant feature of   
 the data as  shown in Figs. 3 and   
4.  Since $|S|^2 \ll |P_\pm|^2$ the $S$-wave is not   
 expected to be significant in the mass spectrum {\it i.e.} in the $\langle   
 Y^0_{\;0} \rangle$ moment. It will primarily contribute to the   
$\cos\theta$ asymmetry measured by the $\langle Y^1_{\;0} \rangle$ and   
 the  $\langle Y^1_{\;1} \rangle$ moments. For diffractive $P$-wave  production   
 with equal phases of all \mbox{$P$- amplitudes} it is expected that   
 $|\langle Y^1_{\;1} \rangle| > |\langle Y^1_{\;0} \rangle|$   
 since the latter describes the interference between the $S$- and the   
 helicity flip $P_0$-wave which has a smaller magnitude than $P_+$ one.    
 The data suggests, however, that the two   
 moments are comparable with more structure actually seen in the   
 $\langle Y^1_{\;0} \rangle$ moment.   
This can only be possible if we allow  
 a well defined pattern of phases in the three waves. This justifies  
 our choice  of additional parameters as already described above.   
In the following section we discuss the results of fitting this model   
to the experimental data.    
   
\section{ Numerical results}    
   
\hspace{0.5cm}In the analysis of the $S$- and $P$-wave  production we compare the    
 model described above to the $t$, $\M$ and angular distribution of   
 the $\KK$ system at two photon energies, $E_\gamma = 4\mbox{   
 GeV}$~\cite{Barber} and $E_\gamma = 5.65 \mbox{ GeV}$ ~\cite{Fries,Behrend}.  
  The two photon energies represent averages of the photon beam  
 energies used in the experiments performed at Daresbury and at DESY,   
 respectively.  
 At first we discuss the momentum transfer dependence of the cross section   
 integrated over the $\KK$ effective mass and the kaon emission angles.  
   
\subsection{ Momentum transfer distributions and integrated cross sections at   
${\bf E_\gamma = 4\mbox{ GeV}}$ }  
\hspace{0.5cm} At 4 GeV photon energy the comparison between the experimentally measured   
 differential cross section $d\sigma/dt$ and the model cross section  
 integrated over the $\KK$ mass \mbox{$ 0.997$~GeV $< \M < 1.042$~GeV} is  
  shown in \mbox{Fig. 2.}  
\begin{figure}[htbp]  
\begin{center}     
\mbox{\epsfxsize 7cm\epsfysize 11cm\epsfbox{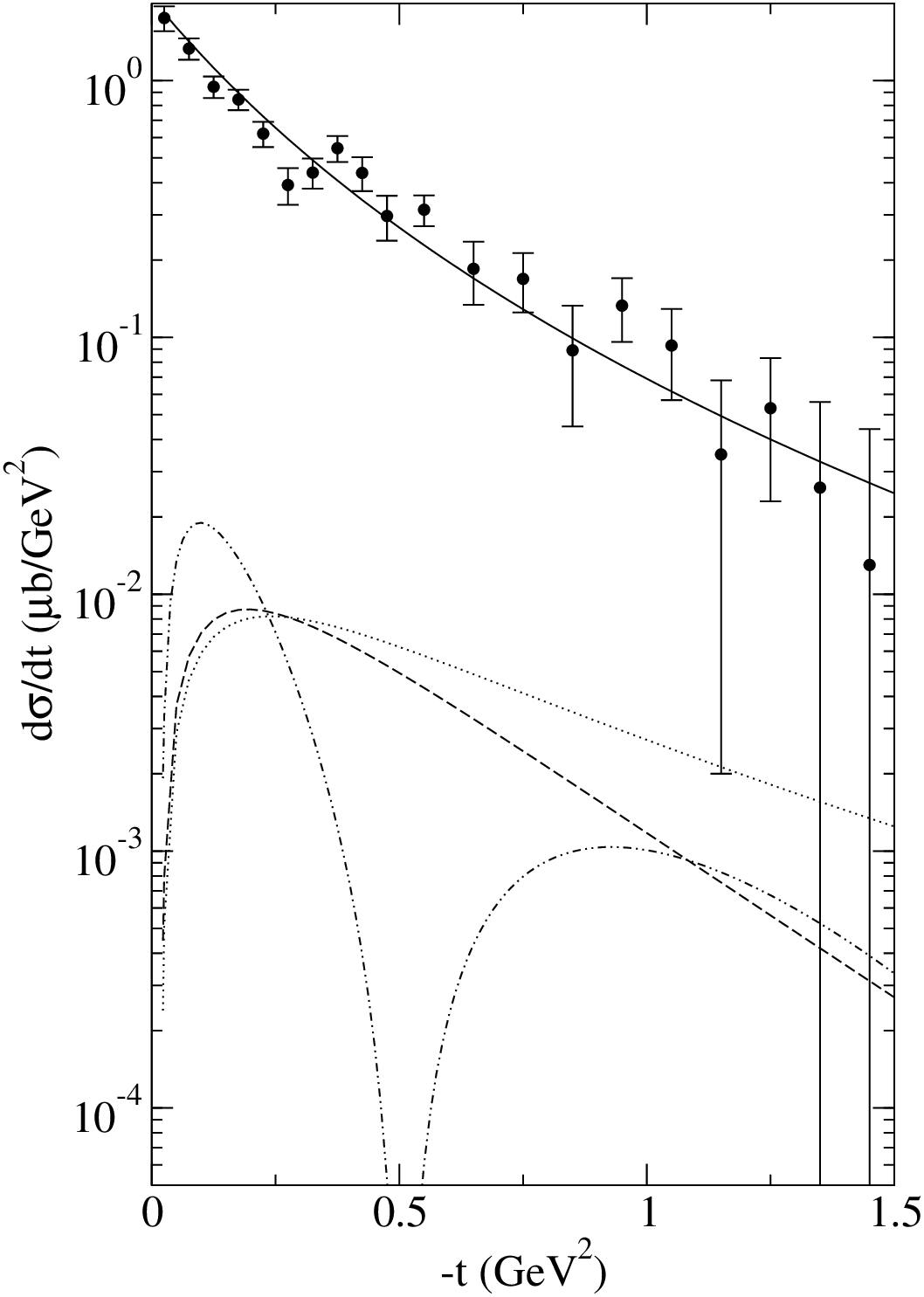}}  
\end{center}     
{Fig. 2: Differential cross section at $E_\gamma= 4$~GeV.   
The solid line shows the model \mbox{$t$-distribution} for the $\phi$ photoproduction,   
the dotted line is the $P$-wave  contribution with M=0   
multiplied by the branching ratio of the $\phi$ decay  
into the $K^+K^-$ pair. The dashed line is the $S$-wave  part of the $K^+K^-$  
cross section calculated for normal $\rho, \omega$ propagators, while the   
double dotted-dashed line corresponds to the Regge propagators. Model  
parameters are given in \mbox{Table 1.} Data are from  
 Fig. 6b of ~\cite{Barber}.}  
\end{figure}     
For this comparison, the $t$-dependent normalization $F(t)$ in the $P$-wave     
 was chosen as    
 \begin{equation}  
 F(t)= \frac{D_1e^{bt}}{(1-t/a)^2} ,                  \label{F}  
 \end{equation}  
 where the normalization constant $D_1$ has been adjusted to reproduce  
 the very forward value of   
 $d\sigma/dt~(t_0)= 1.852~ \mu b/$GeV$^2$ at small argument,   
 $t_0=-0.0225~$GeV$^2$,   
 and the remaining parameters have been chosen as $a=0.7$ GeV$^2$ and   
 $b=0.05$ GeV$^{-2}$. The value of  $d\sigma/dt~(t_0)$ has been  
 obtained from the  
 experimental fits of Barber {\it et al.} at low momentum transfers  
 and in the energy   
 range between 3.4 and 4.8 GeV (see Fig. 7 of ~\cite{Barber}). We took   
 $ d\sigma/dt=(2.13\pm0.38) \mu b/$GeV$^2$~$exp[~(6.2\pm1.3) $GeV$^{-2}~t]$  
  and calculated its value at the minimum $|t_0|$ argument   
 corresponding to the $\KK$ effective mass 1.042 GeV, which was the upper  
 limit studied in \cite{Barber}. Then the constant   
 $D_1$ equals to  
 \begin{equation}  
 D_1= \left(\frac{d\sigma/dt~(t_0)~ BR}{Int}\right)^{\frac{1}{2}} ,  \label{D1}  
 \end{equation}  
 where BR=0.486 is the branching ratio for the $\phi$ decay into $K^+K^-$  
 used in ~\cite{Barber} and   
\begin{equation}  
 Int=\int^{M_2}_{M_1} \frac{d\sigma^P}{dt~d\M}(D_1=1,t_0)~ d\M,       \label{Int}  
 \end{equation}  
 $M_1$ being the lower effective mass limit 0.997 GeV and $M_2$ being   
 the upper   
 mass limit 1.042 GeV. In the above equation the unnormalized double   
 differential cross section $d\sigma^P/(dt~d\M)$  
 corresponds to the $P$-wave  amplitudes calculated at fixed $t_0$.   
 A variation of the $t_0$ with the effective mass $K^+K^-$ has been   
 neglected in this narrow band of $\M$.   
  
 As expected, the $t$-distribution is dominated by the   
 helicity non-flip \mbox{$P$-wa}\-ve. The $P_0$-wave  and the $S$-wave  are   
 kinematically suppressed at low $t$ and are two orders of magnitude smaller   
 than the dominant wave.  
 After integration over $t$ in the range up to $-t=1.5~$GeV$^2$ the total   
 $\phi$ photoproduction cross section equals 0.449 $\mu b$ and is in very  
  good  
 agreement with the measured value of $(0.450\pm0.019)$ $\mu b$ presented in   
 Table 1 of \cite{Barber} for the photon energy range between 2.8 and 4.8 GeV.  
The $K^+K^-$ part of the integrated $P$-wave  cross section equals   
$(0.218\pm0.039)$ $\mu b$. In the model fits to data, the decomposition of the total $P$-wave   
cross section in its parts corresponding to $M=1$, $M=0$ and $M=-1$ components  
depends to some extent on the contribution of the $S$-wave  cross section and   
in particular on the choice of the type of propagators included in the $S$-wave amplitude. For normal $\rho, \omega$ propagators the integrated cross   
sections for the $P_0$-wave  and the $S$-wave  are equal to   
 $(6.4^{+5.5}_{-4.8})$ nb and  $(4.9^{+5.8}_{-3.6})$ nb,  
respectively. The corresponding numbers for the Regge propagators are   
$(4.7^{+5.7}_{-4.5})$ nb and \mbox{$(4.3^{+6.6}_{-3.6})$ nb.} The errors of the cross  
sections have been evaluated using the limits of the model parameters obtained  
from the fitting program MINUIT \cite{MINUIT}. These numbers are smaller  
than our early estimates of the $S$-wave  integrated cross sections written  
in \cite{Ji}. One reason of this change is related to a presence of the  
$S$-wave  form factor $exp(B_St)$ introduced in the previous section and the  
second one comes from the diminution of the $S$-wave  modulus obtained in the  
fitting procedure which will be explained later.  
   
\subsection{ Momentum transfer distributions and integrated cross sections at   
${\bf E_\gamma = 5.65\mbox{ GeV}}$ }  
   
\hspace{0.5cm} The DESY data \cite{Behrend} have been taken at much lower momentum transfers   
 than the Daresbury data \cite{Barber}. The most precise data of Behrend et al.  
 lie within the range $|t-t_{min}|<0.2$~GeV$^2$. The average energy   
 \mbox{$E_{\gamma}=5.65$~GeV} corresponds to the photon energy range between 4.6 
  and 6.7 GeV. The   
 $\phi$ production differential cross section $d\sigma/dt$ can be fitted as   
\mbox{$d\sigma/dt =n ~exp(b t)$}, where $n=(2.40\pm0.15)~\mu b/$GeV$^2$ and  
$b= (6.11\pm 0.53)$ GeV$^{-2}$ represents  
the average slope of $d\sigma/dt$ for the four upper energy bins given in   
Table 3 of \cite{Behrend}. Taking the appropriate $t_0$ value equal to   
$-0.0113 ~ $GeV$^2$  
we calculate $d\sigma/dt~(t_0)=2.24~ \mu b/$GeV$^2$. Then we define a simple  
form of the $t$-dependent normalization factor at $E_{\gamma}=5.65$~GeV as  
 \begin{equation}  
 F(t)= D_2 ~e^{\frac{1}{2} bt} ,                  \label{F2}  
 \end{equation}  
and use once again (\ref{D1}) and (\ref{Int}) to find a new constant $D_2$   
taking into account  
that the $\phi$ branching ratio used in \cite{Behrend} was BR=1/2.14,   
\mbox{$M_1=1.01$ GeV} and $M_2=1.03$ GeV.  
The $K^+K^-$ photoproduction cross section at $E_\gamma =5.65$ GeV   
integrated over $|t|$ up to 0.2 GeV$^2$ is 120.5 nb. Its $P_0$- and  
$S$-wave  parts are $(13.8^{+5.3}_{-4.7})$ nb and  
$(7.0^{+6.8}_{-4.4})$ nb for normal propagators and  
$(14.0^{+5.3}_{-4.8})$ nb and $(6.8^{+6.6}_{-4.3})$ nb for Regge  propagators,  
 respectively. Let us notice that the $P_0$- and $S$-wave  cross sections are  
  comparable and do not vary too much with energy bearing in mind rather large  
  errors. The $S$-wave  cross section at 5.65 GeV is comparable in its  
  magnitude with the estimate of the upper limit $(2.7\pm1.5)$ nb quoted  
in \cite{Fries} and \cite{Behrend}.   
We finally note that the total $\phi$ photoproduction cross  
 section of $(0.25 \pm 0.2)$ $\mu b$ given in~\cite{Fries} corresponds  
 to $M_{\KK}$ integrated over the range 1.0 GeV to 1.024  
  GeV. When integrated in this mass range our model gives  
0.23 $ \mu b$ in good agreement with the measurement.  
   
\subsection{ ${\bf K^+K^-}$ mass distributions and moments at\\    
${\bf E_\gamma = 4\mbox{ GeV}}$ }  
  
\hspace{0.5cm}Next let us describe calculations of the $K^+K^-$ differential cross section  
 integrated over a certain range of the momentum transfer at fixed $\M$ mass.   
 We first discuss the results corresponding to $E_\gamma = 4\mbox{   
 GeV}$ where the upper limit of $-t$ was 1.5 GeV$^2$.  
\begin{table}[ht]  
\caption{Fitted values of model parameters for $E_{\gamma}=4$ GeV.  
Units of $A$ and $B$ are $\mu b/$GeV and $\mu b/$GeV$^2$, respectively.}  
\begin{center}  
\begin{tabular}{|c|c|c|}  
\hline  
 &Normal propagators & Regge propagators \\  
\hline  
$\phi_{00}$ & $122.3^{\rm o} \,_{-21.5^{\rm o}}^{+22.6^{\rm o}}$&$74.5^{\rm o} \,   
_{-27.0^{\rm o}}^{+29.7^{\rm o}}$\\  
\hline  
$\phi_{10}$ & $87.6^{\rm o}  \,_{-11.1^{\rm o}}^{+9.9^{\rm o}}$&$86.8^{\rm o} \,   
_{-23.1^{\rm o}}^{+12.8^{\rm o}}$\\  
\hline  
$C_{00}$ & $0.33 \pm 0.16$ & $0.72 \, _{-0.43}^{+0.42}$\\  
\hline  
$C_{10}$ & $0.44 \,_{-0.22}^{+0.16}$ & $0.37 \, _{-0.31}^{+0.18}$\\  
\hline  
$A$ & $6.65 \, _{-0.23}^{+0.22}$ & $6.67 \, _{-0.24}^{+0.22}$\\  
\hline  
$B$ & $133.0 \pm 11.9 $ & $133.1 \pm 11.9$\\  
\hline  
$v_{10}$ & $(-12.2 \, _{-6.7}^{+6.6})\!\times \! 10^{-3}$& $(-11.3 \pm 6.5)\! \times \! 10^{-3}$\\  
\hline  
$v_{11}$ & $(-2.0 \pm 5.5)\!\times \! 10^{-3}$& $(-1.2 \, _{-5.6}^{+5.5})\! \times \! 10^{-3}$\\  
\hline  
$v_{20}$ & $(-7.8 \, _{-9.0}^{+8.9})\! \times \! 10^{-3}$& $(-6.5 \, _{-9.1}^{+8.9})\! \times \! 10^{-3}$\\  
\hline  
\end{tabular}  
\end{center}  
\end{table}  
\begin{figure}[htbp]  
\begin{center}     
\mbox{\epsfxsize 12.6cm\epsfysize 9.1cm\epsfbox{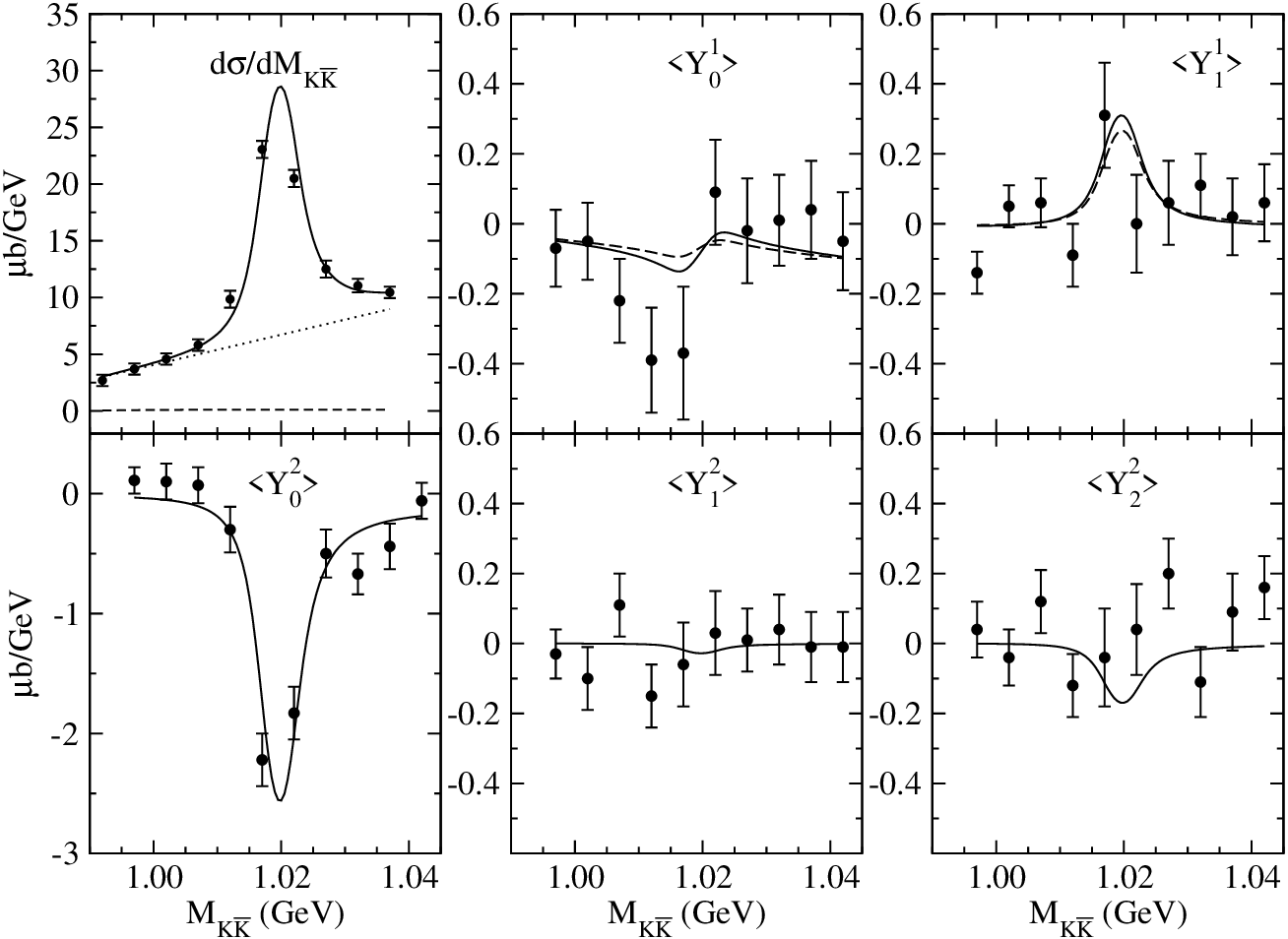}}  
\end{center}     
{Fig. 3: $K^+K^-$ mass spectrum and moments of angular   
distribution in the   
helicity frame for an incident photon energy 4 GeV. Solid lines are results  
of model calculations for the normal $\rho, \omega$ propagators, while the  
dashed lines correspond to the Regge propagators.  
The dotted line shows the background contribution to the mass  
spectrum, while the short dashed line represents the $S$-wave part. The phenomenological   
parameters are given in Table 1. The data points are from ~\cite{Barber}.}  
\end{figure}  
  In Fig. 3 we show the results of the   
 simultaneous fit to the $K^+K^-$ effective mass distribution $d\sigma/d\M$  
in the mass range $0.992 < \M < 1.037\mbox{ GeV}$ and to the moments   
$\langle Y^{L}_{M} \rangle$  at \mbox{ 0.997 GeV~$< \M <$~ 1.042 GeV.}   
 To account for a possibly large $\pi\pi$ experimental background in   
 ~\cite{Barber} we have introduced an additional linear term   
 \begin{equation}  
\frac{d\sigma_{b}}{dM_{K\overline{K}}}=A + B (M_{K\overline{K}}-M_{av}),   
                                                            \label{bg}  
 \end{equation}  
where $A$ and $B$ are free parameters and $M_{av}=(M_1+M_2)/2$ is the average  
$\KK$ effective mass in the range chosen above.  
The parameters $A$ and $B$ will be fitted to the experimental data. The  
background cross section integrated over the mass range between $M_1$ and  
$M_2$ is equal to $A(M_2-M_1)$. It is rather large, attaining a value of about  
300 nb. Similarly we have added the background terms to three moments:  
\begin{equation}  
\langle Y^{1}_{\;0} \rangle_b=v_{10}\frac{d\sigma_{b}}{dM_{K\overline{K}}},~~~~  
\langle Y^{1}_{\;1} \rangle_b=v_{11} \frac{d\sigma_{b}}{dM_{K\overline{K}}},~~~~  
\langle Y^{2}_{\;0} \rangle_b=v_{20}\frac{d\sigma_{b}}{dM_{K\overline{K}}},  
 \label{bgY10,11,20}  
\end{equation}  
where $v_{10}$, $v_{11}$ and $v_{20}$ are constants.  
The remaining two moments $\langle Y^{2}_{\;1} \rangle$ and   
$\langle Y^{2}_{\;2} \rangle$ have not been corrected for background since their   
experimental values fluctuate around zero \cite{Barber}.   
In the fits to data a finite experimental resolution of the effective  
$K^+K^-$ mass distribution was taken into account by choosing the effective  
$\phi$ width equal to 5.6 MeV according to the $K^+K^-$ spectrum in Fig. 5  
of \cite{Barber}. The theoretical effective mass distribution and the  
moments have been smeared over the mass interval $\Delta \M=5$~ MeV   
equal to the mass bin size. Both effects lead to broadening  
of the $\phi$ resonance spectrum and to a widening of the structure of   
 moments near   
the $\phi$ mass.   
The overall number of free parameters in our fit is 9     
 (or 7 if $v_{11}$ and $v_{20}$ are set to zero, which they are within errors).  
 Table 1 presents the values of fitted model parameters in cases of normal   
 and Regge propagators used in the $S$-wave  amplitudes.   
The total number of experimental data is 60. We see in Fig. 3  
 that all the moments are well fitted including the moment   
$\langle Y^1_{\;0} \rangle$. One has the impression that the  
 theoretical curve corresponding to this moment has a slightly too small   
 amplitude  near 1.015 GeV, but the  
  $\chi^2$ value for 10 data points is good: equal to 8.   
 The $S$-wave  cross section  
is small, almost invisible in comparison with the large peak corresponding  
to the $\phi$ resonance and the very important background which we attribute  
to an experimental missidentification of the $\pi\pi$ events as $\KK$  
events. In our fits of the moments we have not assumed like in \cite{Barber} that   
the nucleon non-flip $S$-wave  amplitudes vanish. In fact, they do not vanish  
and become more important at higher values of $t$.  Thus we have included the  
moment $\langle Y^1_{\;1} \rangle$ in our analysis without making the ad hoc   
assumption that it is zero from the beginning. We see in Fig. 3 that  
$\langle Y^1_{\;1} \rangle$ is in general   
non zero and has some structure near 1.02 GeV related to the position of the   
maximum of the dominant $P$-wave . The experimental values of the moment   
$\langle Y^2_{\;1} \rangle$ are particularly small. In our model this   
moment is also small due to a large phase difference $\phi_{10}$ between the     
$P$-wave  amplitudes with $M=1$ and $M=0$. This phase is close to 90$^{\rm o}$  
as seen in Table 1. The $S$-wave phase has also a large correction $\phi_{00}$,  
which depends on the type of propagators used in the model. This phase is  
smaller for the Regge propagators since they are complex and vary with the  
momentum transfer. On the average the Regge propagators add about 50$^{\rm o}$  
to the phase of the \mbox{$S$-wave .} This increase is compensated by a  
\mbox{decrease of $\phi_{00}$.} 
 
 Among the \mbox{$S$-wave} amplitudes,   
the proton spin non-flip components are the most important ones, although their dominance is not so   
strong as in the case of the $P$-wave . This feature of the $S$-wave  amplitudes  
is related to an important contribution of the $\rho$ exchange. The phase   
difference between the $S$-wave   
proton spin non-flip and the $P_0$-wave proton spin non-flip amplitudes is  
larger than 90$^{\rm o}$ for the $\M$ masses smaller than the $\phi$ mass and it   
becomes smaller  
than 90$^{\rm o}$ on the right hand side of the $\phi$ resonance. This happens due  
to the rapid phase increase of the resonant $P_0$-amplitude. As a consequence  
of this phase variation the moment $\langle Y^1_{\;0} \rangle$ has a minimum  
to the left of $\phi$ resonance position and the maximum to its  
right. Let us stress here that only  
this moment has been analyzed in \cite{Barber} as a source of the $S-P$ wave  
interference. We should remark, however that the   
moment $\langle Y^1_{\;1} \rangle$ also depends sensitively on the $S$-wave amplitudes.  
  
In addition to phases the fitting program provides us with the values of the  
moduli $C_{00}$ and $C_{10}$. At 4 GeV both factors are smaller than one  
so the integrated cross sections for the $S$-wave  and the $P_0$-wave are reduced in   
magnitude and their final numbers stay below 10 nb as already written.   
  
\subsection{ ${\bf K^+K^-}$ mass distributions and moments at \\  
${\bf E_\gamma = 5.65\mbox{ GeV}}$ }  
  
\hspace{0.5cm}We pass to a discussion of the angular momentum structure at the average  
photon energy 5.65 GeV. The authors of  ~\cite{Behrend} have presented  
in their \mbox{Fig. 22} the so-called normalized moments of spherical harmonics   
$\langle Y^{L}_{M} \rangle/\langle Y^{0}_{\;0} \rangle$ together with   
the unnormalized  
$\KK$ mass distribution representing numbers of events per 10 MeV bin.  
We have attempted to make our own normalization of the above data in order to   
obtain the functional dependence of the moment $\langle Y^0_{\;0} \rangle$   
which is equal to the differential cross section $d\sigma/d\M$ divided by   
$(4\pi)^{1/2}$. We have integrated the $\phi$ production differential cross   
section $d\sigma/dt$ parameterized above as a simple exponential function in the  
$-t$ range up to 0.2 GeV$^2$ obtaining the value of $(0.258\pm0.020)$ $\mu \rm b$.  
 This value  
corresponds to the sum of $3927\pm82$ $K^+K^-$ events in the $\M$  
range between 1.01 and 1.03 GeV. Taking into account the $K^+K^-$  
branching ratio reported as  
 1/2.14 in ~\cite{Behrend} we get the normalization constant equal to   
$(3.068\pm 0.245) \cdot 10^{-5} \mu b$/event. Knowing this constant one can   
calculate the moment   
$\langle Y^{0}_{\;0} \rangle$ and consequently the values of all other moments   
$\langle Y^{L}_{M} \rangle$ for L and M  up to 2. Then we have performed a   
common fit to   
$d\sigma/d\M$ and 5 moments $\langle Y^{L}_{M} \rangle$ for $\M$ between 1.005 and  
1.045 GeV, including fully the range of the $\phi(1020)$ resonance as well as  
a part of $\M$ well above it. Here the width of the mass bins was 10 MeV. 
Unfortunately the value of the mass distribution $d\sigma/dM_{K\bar K}$ 
corresponding to the extreme experimental data point at 0.995 GeV was not given 
in ~\cite{Behrend}, even though the $L\ne 0$ normalized moments were presented. 
For this reson we were unable to include this bin in our fit. 
The effective mass resolution reported in ~\cite{Behrend} was about 7 MeV  
and the effective $\phi$ width chosen in the analysis done by ~\cite{Fries}   
was 8 MeV. We included the finite mass resolution by smearing our theoretical   
mass spectrum and moments over the 8.5 MeV range around each $\M$ value.  
In the \mbox{$P$-wave } amplitudes the Breit-Wigner form of the $\phi$ spectrum was  
 used with the $\phi$ mass equal to 1.0194 GeV and the width equal to 4.26 MeV.   
\begin{figure}[htbp]  
\begin{center}     
\mbox{\epsfxsize 12.6cm\epsfysize 9.1cm\epsfbox{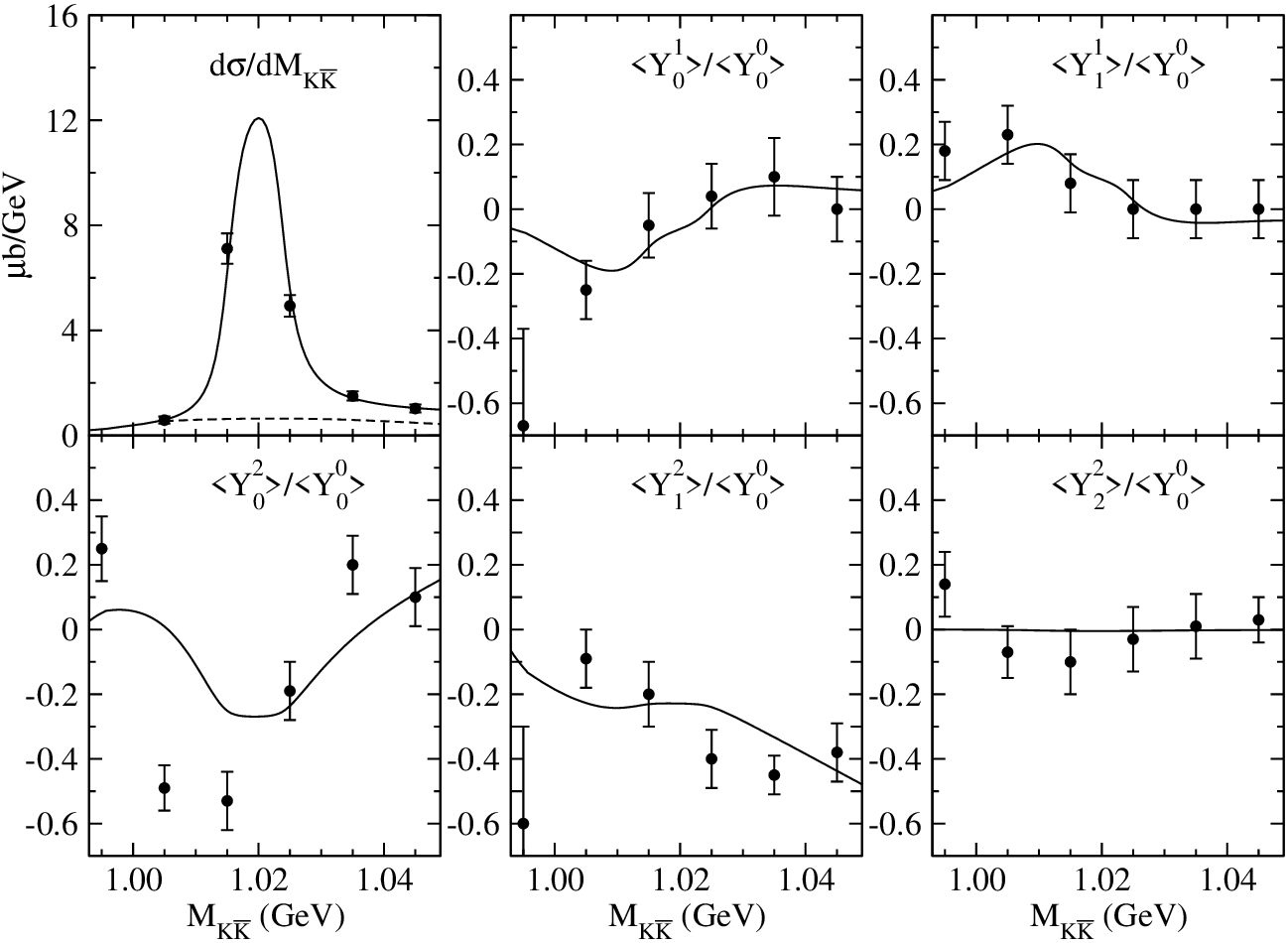}}  
\end{center}     
{Fig. 4: $K^+K^-$ mass distribution and normalized moments  
of angular   
distribution in the helicity system for an incident photon energy 5.65 GeV.   
The solid lines are results  
of model calculations; the dashed line is the $S$-wave contribution to the mass  
distribution. The phenomenological parameters are given  
in Table 3. The data points are from \cite{Behrend}.}  
\end{figure}  
\begin{table}[ht]  
\caption{The fitted values of model parameters for $E_{\gamma}=5.65$ GeV.  
 Units of $A$ are $\mu b/$GeV and $B$, $\beta_{20}$ and $\beta_{21}$ are in  
$\mu b/$GeV$^2$.}  
\begin{center}  
\begin{tabular}{|c|c|c|}  
\hline  
 & Normal propagators & Regge propagators\\  
\hline  
$\phi_{00}$ & $106.0^{\rm o} \, _{-16.2^{\rm o}}^{+10.2^{\rm o}}$ & $49.3^{\rm o}  
 \, _{-16.0^{\rm o}}^{+10.1^{\rm o}}$\\  
\hline  
$\phi_{10}$ & $11.4^{\rm o} \, _{-16.6^{\rm o}}^{+17.3^{\rm o}}$&  $13.0^{\rm o} \,  
 _{-16.7^{\rm o}}^{+17.2^{\rm o}}$\\  
\hline  
$C_{00}$ & $1.06 \, _{-0.41}^{+0.43}$ & $1.53 \, _{-0.59}^{+0.61}$\\  
\hline  
$C_{10}$ & $1.59 \, _{-0.30}^{+0.28}$ & $1.60 \, _{-0.30}^{+0.28}$\\  
\hline  
$A$ & $0.23 \, _{-0.31}^{+0.21}$ & $0.24 \, _{-0.29}^{+0.21}$\\  
\hline  
$B$ & $7.14 \, _{-4.80}^{+4.81} $ & $7.22 \, _{-4.80}^{+4.82}$\\  
\hline  
$\beta_{20}$ & $1.20 \pm 0.45$ & $1.19 \pm 0.45$\\  
\hline  
$\beta_{21}$ & $-1.98 \pm 0.45$ & $-1.98 \pm 0.45$\\  
\hline  
\end{tabular}  
\end{center}  
\end{table}  
As at 4 GeV energy we have introduced a linear background term in the   
effective $K^+K^-$ mass distribution and in the two moments:  
\begin{equation}  
\langle Y^{2}_{~0} \rangle_b=\beta_{20}(M_{KK}-M_{th}),~~~~  
\langle Y^{2}_{~1} \rangle_b=\beta_{21}(M_{KK}-M_{th}),       \label{Y20,Y21b}  
\end{equation}  
where $\beta_{20}$ and $\beta_{21}$ are parameters, and $M_{th}$ is the   
threshold mass of the $K\overline{K}$ system. Counting the four parameters  
in the two complex factors $C_{00}~exp(i\phi_{00})$ and $C_{10}~exp(i\phi_{10})$   
which  
modify the $S$-wave  and the $P_0$-wave  amplitudes, and adding two   
 background parameters $A$ and $B$, we have altogether   
eight parameters to be fitted to data at 5.65 GeV. The results of the fits are  
shown in Fig. 4 and the parameters are written in Table 2.  
Contrary to the previous case of $E_{\gamma}=4$ GeV, the background cross   
section at 5.65 GeV is much smaller, less than 5 nb. The shape of the $\KK$  
mass spectrum and the general behaviour of the moments are well described by  
the model perhaps except of the two points of $\langle Y^2_{\;0} \rangle $ at 1.005 and 1.015 GeV.  
The values corresponding to these data  
points are {\em smaller} than $-0.45$. Let us  
notice that the {\em lowest} limit of 
 $\langle Y^2_{\;0} \rangle/\langle Y^0_{\;0} \rangle$ equals to  
 $-1/\sqrt5 \approx -0.45$ if one assumes that only $S-$ and $P-$ waves  
 participate 
 in the $K^+K^-$ production process. Strictly speaking, this limit  
 corresponds to  
the case in which the amplitudes $P_+$ dominate near the position of the  
$\phi(1020)$ resonance. Any admixture of the $P_0 -$, $P_- -$ or $S-$ waves must 
increase the value of $\langle Y^2_{\;0} \rangle/\langle Y^0_{\;0} \rangle$  
above $-1/\sqrt 5$. A uniform background contribution would have the same  
effect. Thus, one is tempting to explain the low experimental values  
at 1.005 and 1.015 GeV by the presence of a background coming from higher waves  
like $D-$ or $F-$ waves. Interestingly, slightly above the $\phi(1020)$ mass, 
 there is a structure in the ratio of  
  $\langle Y^4_{\;0} \rangle/\langle Y^0_{\;0} \rangle$ measured in [17]. This 
structure can be attributed  
to $D-$ wave or to an interference of the $P-$ wave with the $F-$ wave. Both 
cases are, however, physically rather improbable, because near the $K\overline{K}$ 
threshold these waves should be strongly suppressed and we do not know any 
$D$ or $F$ resonances located closely to 1 GeV. Let us also remark that 
a general shape of $\langle Y^2_{\;0} \rangle/\langle Y^0_{\;0} \rangle$  
shown by the line in Fig. 4 is correct near the $K\overline{K}$ threshold since 
the extreme experimental point at 0.995 GeV lies above and not too far from 
the curve. In Fig. 4 we do not show a line corresponding to the small background to  
$d\sigma/d\M$, since its magnitude is very close to the $S$-wave    
contribution shown in this figure.  
   
Finally, in Table~\ref{summary} we list the contributions of the   
 individual waves to the $\KK$ photoproduction cross section at the   
 two energies studied.   
  
\begin{table}[ht]  
\caption{Integrated cross sections in nb}  
\label{summary}  
\begin{center}  
\begin{tabular}{|c|c|c|c|c|}  
\hline  
photon energy&\multicolumn{2}{c|}{4 GeV}&\multicolumn{2}{c|}{5.65 GeV}\\  
\hline  
$S$-wave propagator&normal&Regge&normal&Regge\\  
\hline  
sum of all $P$-waves&\multicolumn{2}{c|}{$218.4\pm 39.5$}&\multicolumn{2}{c|}{$120.5\pm 9.4$}\\  
\hline  
$P_{0}$-wave&$6.4_{-4.8}^{+5.5}$&$4.7_{-4.5}^{+5.7}$&$13.8_{-4.7}^{+5.3}$&$14.0_{-4.8}^{+5.3}$\\  
\hline  
$S$-wave&$4.9_{-3.6}^{+5.8}$&$4.3_{-3.6}^{+6.6}$&$7.0_{-4.4}^{+6.8}$&$6.8_{-4.3}^{+6.6}$\\  
\hline  
background&$299.4_{-10.4}^{+10.0}$&$300.0_{-10.7}^{+10.0}$&$4.5_{-6.1}^{+4.3}$&$4.7_{-5.8}^{+4.2}$\\  
\hline  
$|t|_{max}$&\multicolumn{2}{c|}{1.5 GeV$^2$}&\multicolumn{2}{c|}{0.2 GeV$^2$}\\  
\hline  
$M_{K\overline{K}}$ range&\multicolumn{2}{c|}{(0.997,1.042)~GeV}&  
\multicolumn{2}{c|}{(1.01,1.03)~GeV}\\  
\hline  
\end{tabular}  
\end{center}  
\end{table}  
\subsection{Model predictions at the energy upgraded Jefferson Laboratory}  
 
\hspace{0.5cm}We have performed calculations of the $K^+K^-$ mass spectrum and
 moments  
at $E_{\gamma}$=8 GeV which will be a typical energy of the planned upgrade  
of the CEBAF accelerator operating at the Thomas Jefferson Laboratory.  
\begin{figure}[htbp]  
\begin{center}     
\mbox{\epsfxsize 12.6cm\epsfysize 9.1cm\epsfbox{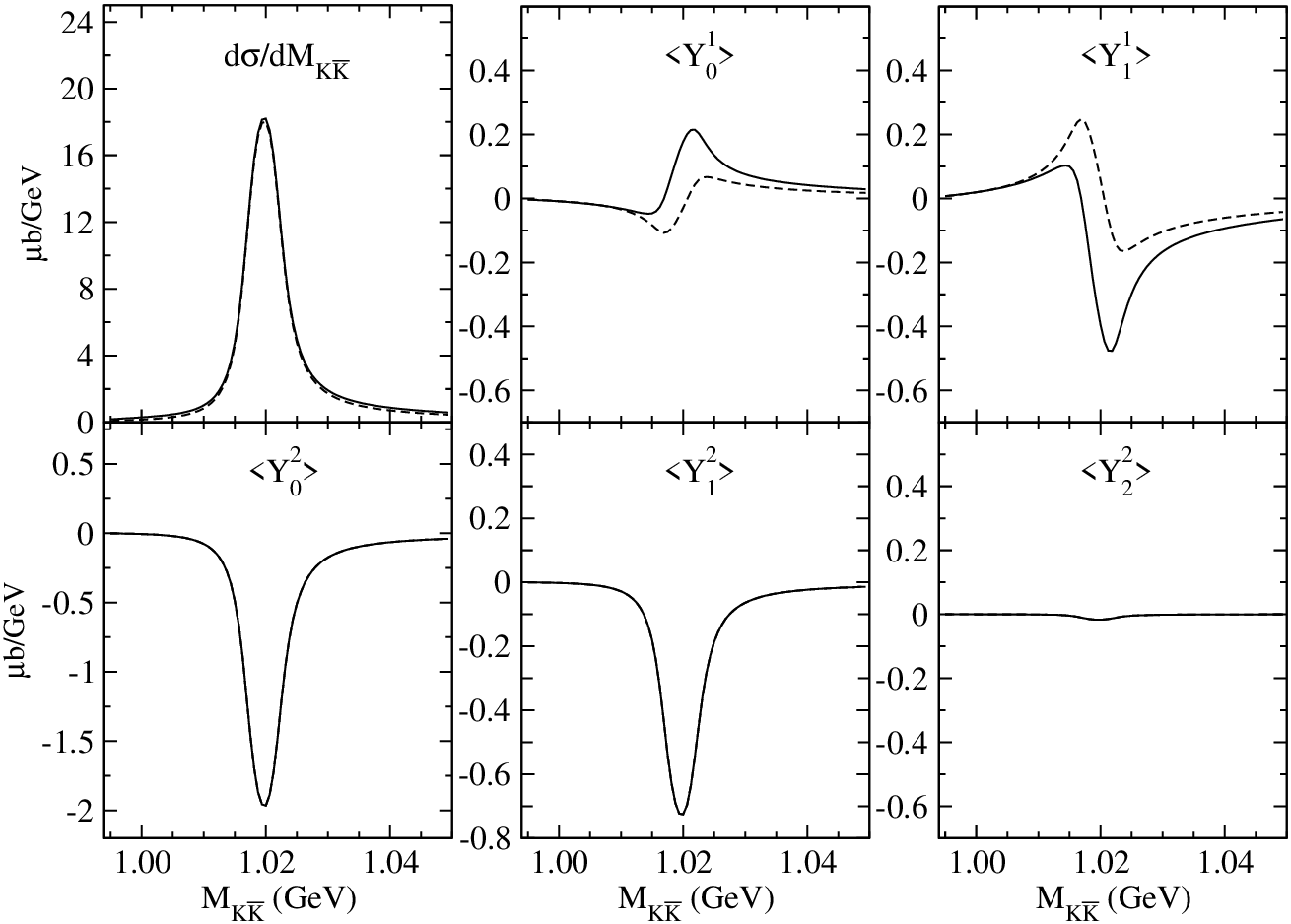}}  
\end{center}     
{Fig. 5: $K^+K^-$ mass spectrum and moments of angular   
distribution in the   
helicity frame for incident photon energy 8 GeV. Solid lines are results  
of model calculations for the normal $\rho, \omega$ propagators while the  
dashed lines correspond to the Regge propagators in the $S-$ wave amplitudes.}  
\end{figure}  
The results presented   
in Fig. 5 can be directly compared with Fig. 3 corresponding to the much lower  
energy of 4 GeV. The calculations have been performed for an ideal case in  
which  
there is no background and no phenomenological adjustment of the moduli and  
phases of the  
$S-$ and $P_{0}-$ waves i.e. $\phi_{00}=\phi_{10}=0$, $C_{00}=C_{10}=1$. We  
have  
assumed that the $K^+K^-$ mass resolution is equal to 5 MeV. The parameters of 
the $\phi(1020)$ meson, like the mass, width and the $K^+K^-$ decay fraction, 
have been taken as 1019.456 MeV, 4.26 MeV and 0.492, respectively. The  
differential 
cross section at low $t$ was parameterized as \mbox{$d\sigma/dt =n ~exp(b t)$} 
with $n= 2.53~ \mu b/$GeV$^2$ and $b= 6.11$ GeV$^{-2}$.  
 One can notice that the  
application of Regge propagators in the $S-$ wave leads to smaller values of  
the $S-$ wave cross section and to smaller amplitudes of the moments  
$\langle Y^{1}_{0} \rangle$  
and $\langle Y^{1}_{1} \rangle$ sensitive to the interference between the  
$S-$ and $P-$ waves. Qualitatively we do not observe important differences in a  
behaviour of the unnormalized moments between the photon energy of 5.65 and 8  
GeV (let us recall here that Fig.4 shows the ratios of  
$\langle Y^L_{\;M} \rangle/\langle Y^0_{\;0} \rangle$,  
not $\langle Y^L_{\;M}\rangle$). 
 
The $S-$ wave total cross sections integrated over the  
$M_{K\overline{K}}$  
range between 1.01 and 1.03 GeV and in the $|t|-$ range from 0.0054 up to  
0.2 GeV$^{2}$ are equal  
to 6.9 nb and 3.0 nb for the normal and Regge propagators, respectively. The  
$P-$ wave  
cross section integrated in the same ranges equals 141 nb while the  
corresponding $P_{0}-$ cross section is equal to 6 nb.  
  
\section{Conclusions}    
\hspace{0.5cm}In this paper we presented results of a theoretical analysis of data on   
 photoproduction of the $K^+K^-$ pairs in the laboratory photon energy range  
 $E_\gamma$ between 2.8 and 6.7 GeV.  In particular we mapped out the  
 interference pattern between the $S$- and the $P$-wave  due to the  
 presence of the scalar $f_0(980)$, $a_0(980)$ and the vector $\phi(1020)$  
 resonances. In the analyses the $S$-wave was described by a  
 model based on the dominance of the  $t$-channel exchange process   
 with the two-meson spectrum described in terms of the  coupled channel   
 meson-meson scattering  $S$- matrix.  The $P$-wave  was described in  
 terms of diffractive production of the $\phi$ resonance  
 decaying into the $K^+K^-$ system. The $S$-wave  and the $P$-wave  contain 4  
 and 12 independent amplitudes respectively and  we included   
 them all while previous analyses made a severe truncation to a  
 single amplitude in each wave.   
 We have shown that amplitudes omitted in the analyses of  
 ~\cite{Fries,Barber} corresponding to the proton  
 helicity non-flip are large and cannot be ignored. In the previous  
 analyses only the $\langle Y_{10} \rangle$ moment was taken into  
 account and this led to a large variation in the estimate of   
 the $S$-wave   photoproduction cross section, between $(2.7\pm 1.5)\mbox{ nb}$  
 at $E_\gamma = 5.65\mbox{ GeV}$~\cite{Fries,Behrend} and $(96.2\pm20)\mbox{  
 nb}$ at $E_\gamma=4\mbox{ GeV}$~\cite{Barber}. In this paper we have  
 considered all six moments appearing in the model with $S$- and $P$-waves  
 which enabled us to isolate the \mbox{$S$-wave } production cross section from  
 that of an incoherent background.   
  This led to a significant, order of magnitude, reduction in the   
 uncertainty in the $S$-wave  production and  reduced the value of  
 cross section at $4\mbox{ GeV}$ from  $96\mbox{ nb}$ to  
 approximately  $5\mbox{ nb}$, thus eliminating the  
 discrepancy between the two measurements (also suggested  
 in~\cite{Barber}). Thus we have found that the $S$-wave  
 photoproduction cross section integrated over the $\phi$ resonance  
 region is between 4 and 7 nb for the two photon energies 4 and 5.65  
 GeV. The cross section of the photon helicity-flip $P_0$-wave,    
  which interferes with the $S$-wave  is found to be somewhat larger  
 than the $S$-wave  cross section, in the range  between 5 and 14 nb.   
 New more precise measurements of the $\KK$ photoproduction with a   
 {\it simultaneous}  
 determination of the $\pi\pi$ cross-section in the $\pi\pi$ effective mass  
  region near 1$\mbox{ GeV}$ could provide new insight into the still  
 controversial nature of the scalar mesons $f_0(980)$ and  
 $a_0(980)$. 
\\  
       
{\bf Acknowledgments} \vspace{0.1 cm}  
  
 We thank Chueng - Ryong Ji and Robert Kami\'nski for their  
 collaboration on the  $S$-wave   photoproduction and for an exchange of   
 ideas on the $P$-wave   amplitudes in the early stage of this work.  
 One of us (L. L.) would like to thank Maria R\'o\.za\'nska and Jacek Turnau for   
 enlightening discussions on the analysis of experimental data.   
 This work was supported in part (A. P. S.) by the US Department of Energy  
 grant under contract   DE-FG0287ER40365.

\end{document}